\documentclass[twocolumn]{aastex6}

\usepackage{natbib}
\usepackage{xcolor}

\newcommand\ts{\thinspace}
\newcommand\E{1E{\thinspace}1740.7$-$2942}
\newcommand{\GG}[1]{}
\begin{document}
\title{
       Tandem Swift and INTEGRAL data to revisit the orbital  \\
       and superorbital periods of 1E{\ts}1740.7$-$2942
      }
\author{
        Paulo Eduardo Stecchini, Manuel Castro,                \\
        Francisco Jablonski, Flavio D'Amico, and Jo\~ao Braga
       }
\affil{
       Instituto Nacional de Pesquisas Espaciais - INPE        \\
       Av. dos Astronautas 1758, 12227-010, S.J.Campos-SP, Brazil \\
      }
\begin{abstract}

  The black hole candidate 1E{\ts}1740.7$-$2942 is one of the strongest 
  hard X-ray sources in the Galactic Center region. No counterparts in 
  longer wavelengths have been identified for this object yet. The 
  presence of characteristic timing signatures in the flux history of 
  X-ray sources has been shown to be an important diagnostic tool for 
  the properties of these systems. Using simultaneous data from 
  NASA's \textit{Swift} and ESA's \textit{INTEGRAL} missions, we have found two 
  periodic signatures at $12.61${\ts}$\pm${\ts}$0.06${\ts}days and 
  $171.1${\ts}$\pm${\ts}$3.0${\ts}days in long-term hard X-ray light 
  curves of 1E{\ts}1740.7$-$2942. We interpret those as 
  the orbital and superorbital periods of the object, respectively. The reported 
  orbital period is in good agreement with previous studies of 
  1E{\ts}1740.7$-$2942 using NASA's \textit{RXTE} data. We present here 
  the first firm evidence of a superorbital period for {\E}, which has
  important implications for the nature of the binary system.
  
\end{abstract}
\section{Introduction} \label{sec:intro}

Since the production of the first hard X-ray (i.e.,
$E${\ts}$>${\ts}20{\ts}keV) image of the Galactic Center (GC) region
by the XRT telescope \citep{1987Natur.330..544S}, {\E}, discovered by
the \textit{Einstein} satellite \citep{1984ApJ...278..137H}, is known to be one
of the strongest X-ray emitters around the GC. With the advent of nearly
continuous sensitive monitoring of the GC region, first provided 
by the {\it SIGMA} telescope \citep{1991ApJ...383L..49S}, {\E} was
found to have spectral states resembling those of Cyg{\ts}X-1.
Then, by analogy, the source has been ever since considered to be 
a black hole candidate (BHC). Its probable black hole nature was further 
supported with the observations of radio jets 
\citep{1992Natur.358..215M} coming from the central X-ray source.

The nature of {\E} system is still a matter of debate in the
literature, with the favored hypothesis being that the system is a
high-mass X-ray binary (HMXB; \cite{2002ApJ...578L.129S}. The deepest
search in the infrared up to date \citep{2010ApJ...721L.126M} opened a
possibility of an extragalactic nature for {\E}, but this has been
recently ruled out by \cite{2015A&A...584A.122L}.

From previous studies \citep[e.g.,][]{cas14}, {\E} is known to spend
most of its time in the low/hard state, with most of the flux in hard
X-rays.  Hence, any attempt in finding a periodic signature in a
long-term light curve can be advantageously carried out with a hard
X-ray database. In this study, we gather data from the {\it Swift}
\citep{2004ApJ...611.1005G} and \textit{INTEGRAL} \citep{2003A&A...411L...1W}
missions to show evidence of two periodic signatures in the flux
history of {\E}, which we interpret as the orbital and superorbital
periods of the source.

\section{Data Selection and Analysis}

The BAT telescope \citep{2005SSRv..120..143B} at the \textit{Swift}
satellite provides daily measurements of the 15--50{\ts}keV flux of
{\E}.  These measurements are not adequate for spectral analysis due
to the sensitivity constraints of a large field imager in X-rays.
However, the ISGRI telescope \citep{2003A&A...411L.141L} on board the
\textit{INTEGRAL} satellite can be used to trace the spectral evolution of
{\E}, as well as to provide the source's flux.  This is where we
tandem the strength of both missions to obtain long-period light
curves of {\E} in its canonical low/hard state (see, e.g.,
\citealp{2005A&A...433..613D}). In such a way any possible signature
in the light curves due to intrinsic spectral variability can be
disentangled from other effects. As both BAT and ISGRI are coded--mask
imaging instruments, any flux contamination due to source confusion is
also avoided, and, thanks to the sensitivity of both telescopes, we
can perform flux measurements at hard X-rays, where {\E} is
brighter. To our knowledge, this is the first time that a search for
periodic signatures in {\E} can be done without source
confusion/contamination and at hard X-rays.

For this work, all the public ISGRI database of {\E} from 2003 to 2015
was retrieved, encompassing 362 spectra. A previous study of this
spectral database by our group was presented by \cite{cas141}. The
data reduction was performed using the OSA software (see, e.g.,
\citealp{2013arXiv1304.1349C}) and a power-law model was found to fit
appropriately each of the 20--200{\ts}keV ISGRI spectra. Any spectrum
with signal-to-noise ratio (SNR) less than 5 and with
${\chi}^{2}_{red}$ (provided by XSPEC) greater than 2 in this energy
range was discarded. The median power-law index was
${\Gamma}${\ts}$=${\ts}$1.8${\ts}$\pm${\ts}0.2. Hence, all
observations with $1.6${\ts}$\leq${\ts}${\Gamma}${\ts}$\leq${\ts}$2.0$
were selected, as this indicates that the source is at its canonical
low/hard state (see, e.g., \citealp{2006ARA&A..44...49R}). Further, we
truncated the ISGRI database to make it begin at the same start date
as the BAT database. As a result we had a total of 162 spectra, from
which the flux in the 20--50{\ts}keV range was used to build a
ISGRI/{\E} long-term 2005--2015 light curve.

The BAT daily flux measurements were chosen to match exactly the same
dates for which a spectrum could be extracted from ISGRI, obeying the
criteria explained above. This guarantees that the source was at its
canonical state, thus allowing us to build a BAT/{\E} long-term light
curve.  For every point removed from the BAT light curve -- based on
the discarded ISGRI points -- the immediate neighbor points were also
discarded (since \textit{INTEGRAL} mission's orbital period is 3 days). Finally,
although the daily coverage of BAT extends further in time, the data
were selected up to 2015 to match the final date in the ISGRI
database. From the 3459 initial measurements, 1599 BAT data
points remained and were used in our analysis.

\section{Results}

\begin{figure}[ht!]
\figurenum{1}
\epsscale{1.13}
\hspace{-0.4cm}
\plotone{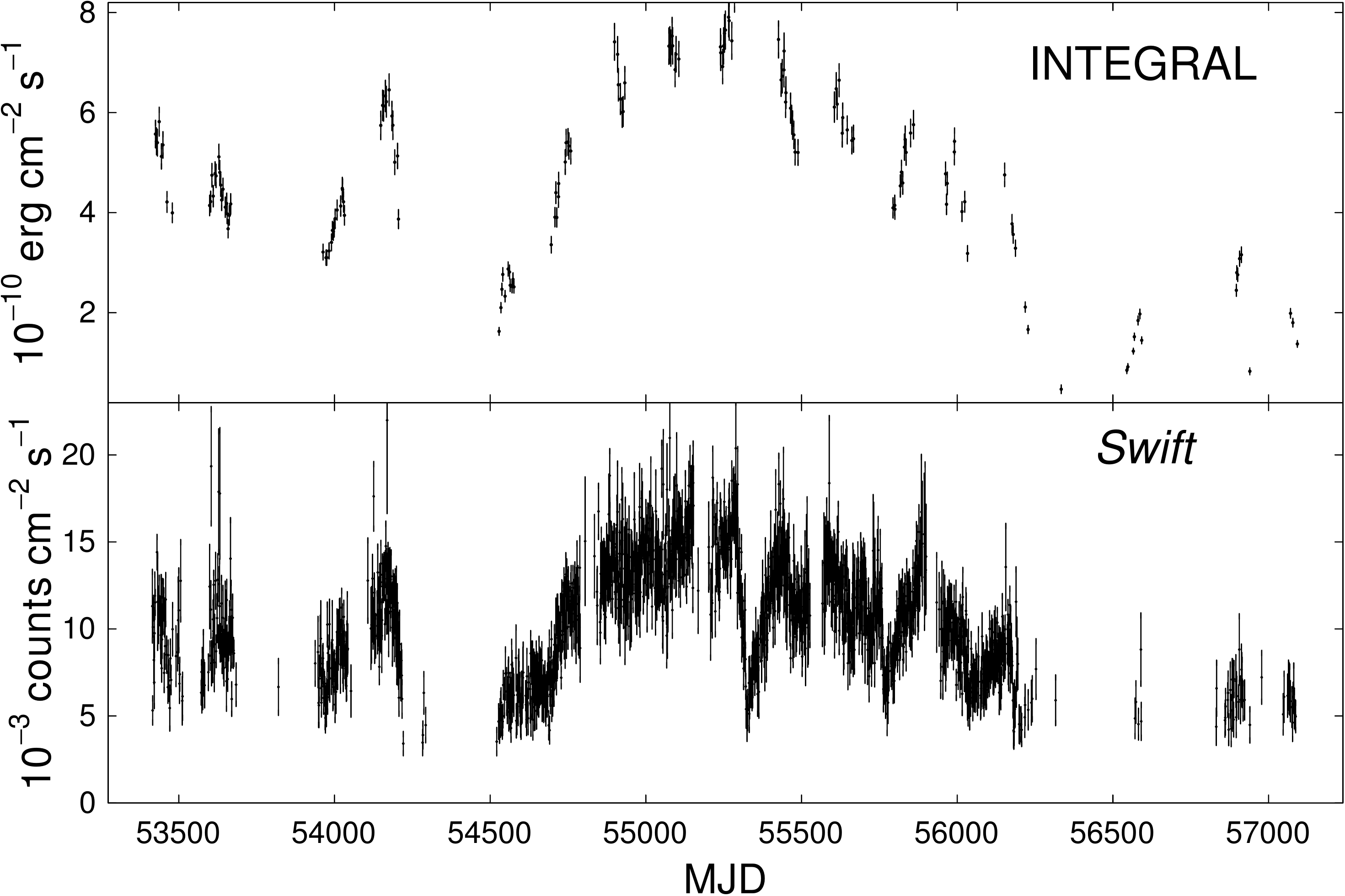}
\caption{
         Long-term light curves of {\E} in its canonical low/hard state, between 2005--2015,
         for the ISGRI and BAT instruments on board \textit{INTEGRAL} and {\it Swift}, respectively.
         \label{fig01}
        }
\end{figure}

The resulting light curves for both \textit{INTEGRAL} and {\it Swift} data,
after our selection criteria and data reduction/analysis, are shown in
Figure{\ts}{\ref{fig01}}.

\begin{figure}[ht!]
\figurenum{2}
\epsscale{1.2}
\hspace{-0.4cm}
\plotone{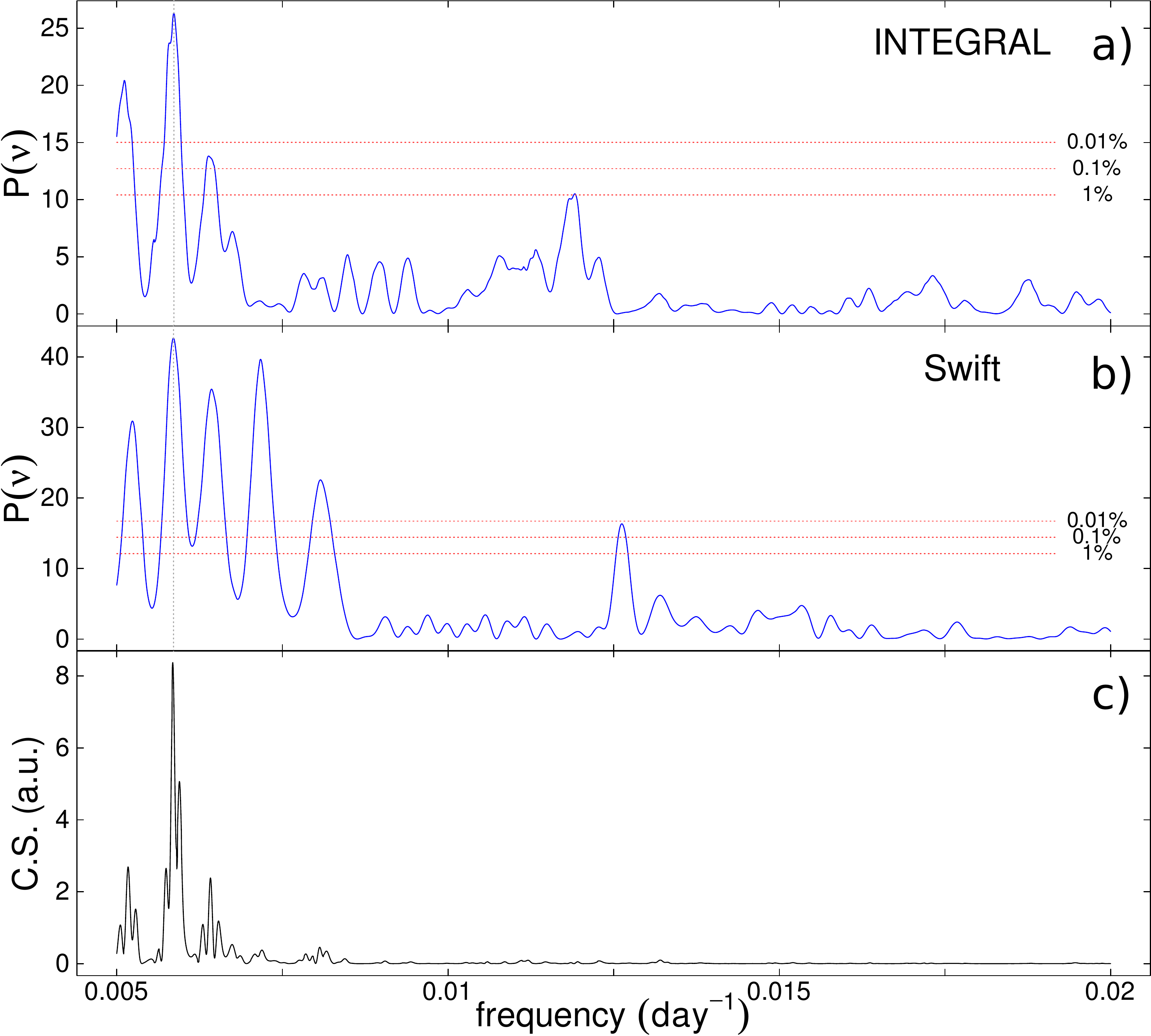}
\caption{
         Lomb--Scargle periodograms for the low-frequency region of {\E}, for both \textit{INTEGRAL} (a) and \textit{Swift} (b) datasets. A strong
         $\sim${\ts}171-day signal is present in the data of
         both missions. False-alarm levels of 1\%, 0.1\% and 0.01\%, calculated according to \cite{1982ApJ...263..835S},
         are also shown (horizontal red dotted lines). In (c), the cross-spectrum of the two independent datasets.
         \label{fig02}
        }
\end{figure}

Data were first explored in the low-frequency
domain. Figure{\ts}{\ref{fig02}} shows, for both \textit{INTEGRAL} (a) and
{\it  Swift} (b), the Lomb--Scargle periodogram from 0.005 to 0.020
days$^{-1}$ (corresponding to 200--50 days), with a grid of 5000
equally spaced frequencies. Low frequencies dominate the periodograms,
and a component with a period around 171 days ($\sim${\ts}0.00585
days$^{-1}$) is present in both databases (vertical gray dotted lines)
with a high level of significance.  A cross-spectral analysis (see,
e.g., \citealp{1989ApJ...343..874S}), shown in
Figure\,{\ref{fig02}}\,(c), identified this period to be
$171.1${\ts}$\pm${\ts}$3.0${\ts}days, with the quoted conservative
uncertainty determined by HWHM of the
cross-spectrum peak. The marked difference between the power spectra
at low frequencies for the \textit{INTEGRAL} and \textit{Swift} datasets is
explained by the better sampling of the latter. This is particularly
noticeable in the light curve, for instance, at MJD $\sim$\,55300, with
a pronounced dip that was sampled only by the BAT instrument. If the
sidelobes seen in the cross-spectrum were related to amplitude
modulation, the corresponding timescale would be $\sim$\,1600\,days --
which is a very prominent feature in the light curves. We interpret
the additional maxima seen in the low-frequency region as due to a
process with P$(\nu)\propto1/\nu$, very common in other X-ray binaries
(see, e.g., \citealp{2008ApJ...675.1424C} and
\citealp{2006ApJS..163..372W}).

\begin{figure}[ht!]
\figurenum{3}
\epsscale{1.13}
\hspace{-0.4cm}
\plotone{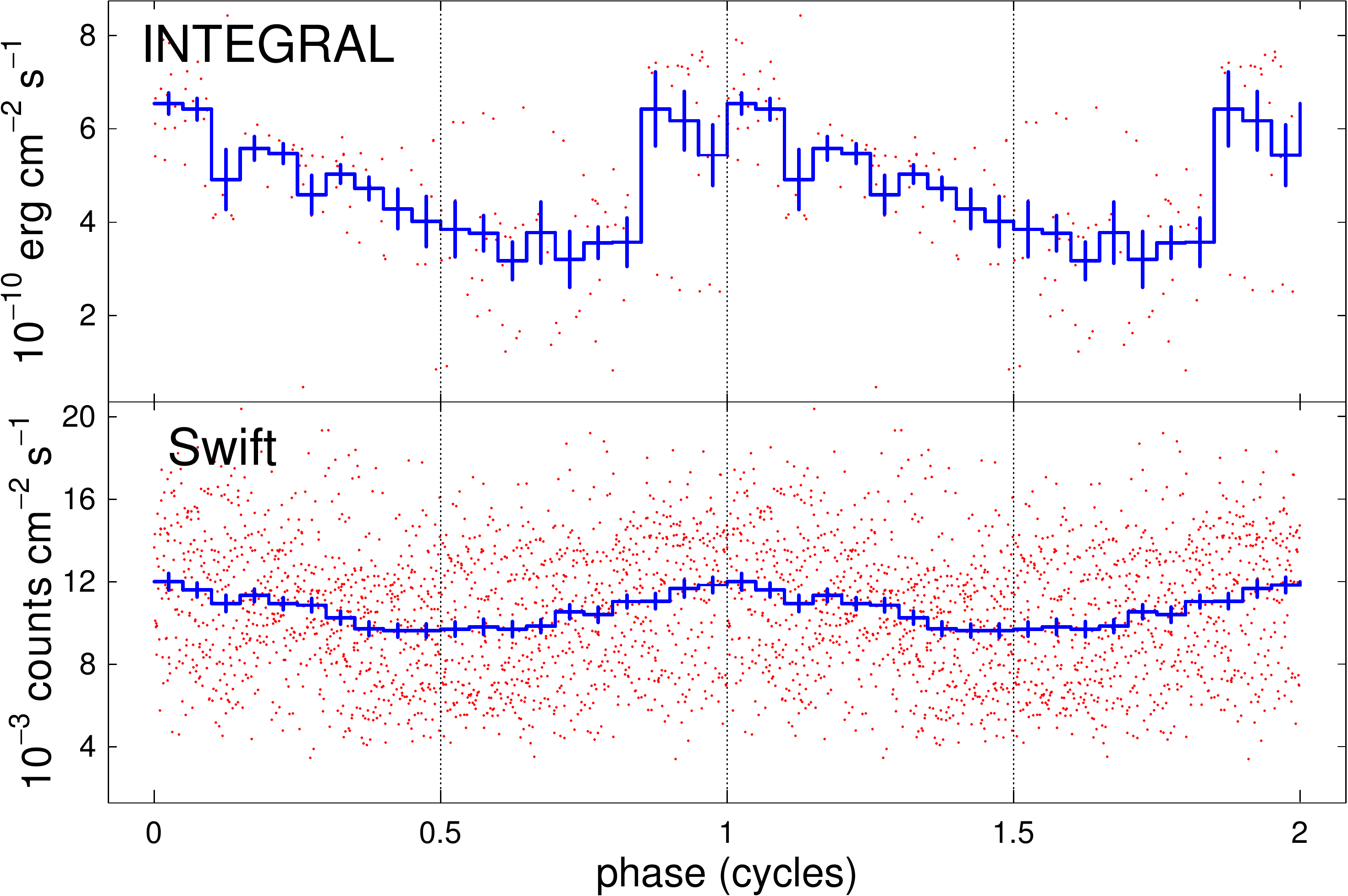}
\caption{
         Folded-phase long-term light curves of {\E} for the superorbital period
         ($171.1${\ts}days) in both our databases. The vertical limits in both
         panels were chosen to provide roughly the same relative range.
         \label{fig03}
        }
\end{figure}

The folded-phase light curve for the times of maximum with the
ephemeris MJD\,53552{\ts}$\pm${\ts}$E$\,$\times${\ts}$171.1${\ts}days,
where $E$ is the number of cycles from the origin, is presented in
Figure{\ts}{\ref{fig03}}, for both sets of data. Although another value
for the superorbital period in {\E} was mentioned before in the
literature \citep{2002ApJ...578L.129S}, we present the first robust
measurement of such modulation, as it is clearly present in the
database of two imaging telescopes -- thus free of source confusion
and flux contamination.

\begin{figure}[ht!]
\figurenum{4}
\epsscale{1.13}
\hspace{-0.55cm}
\plotone{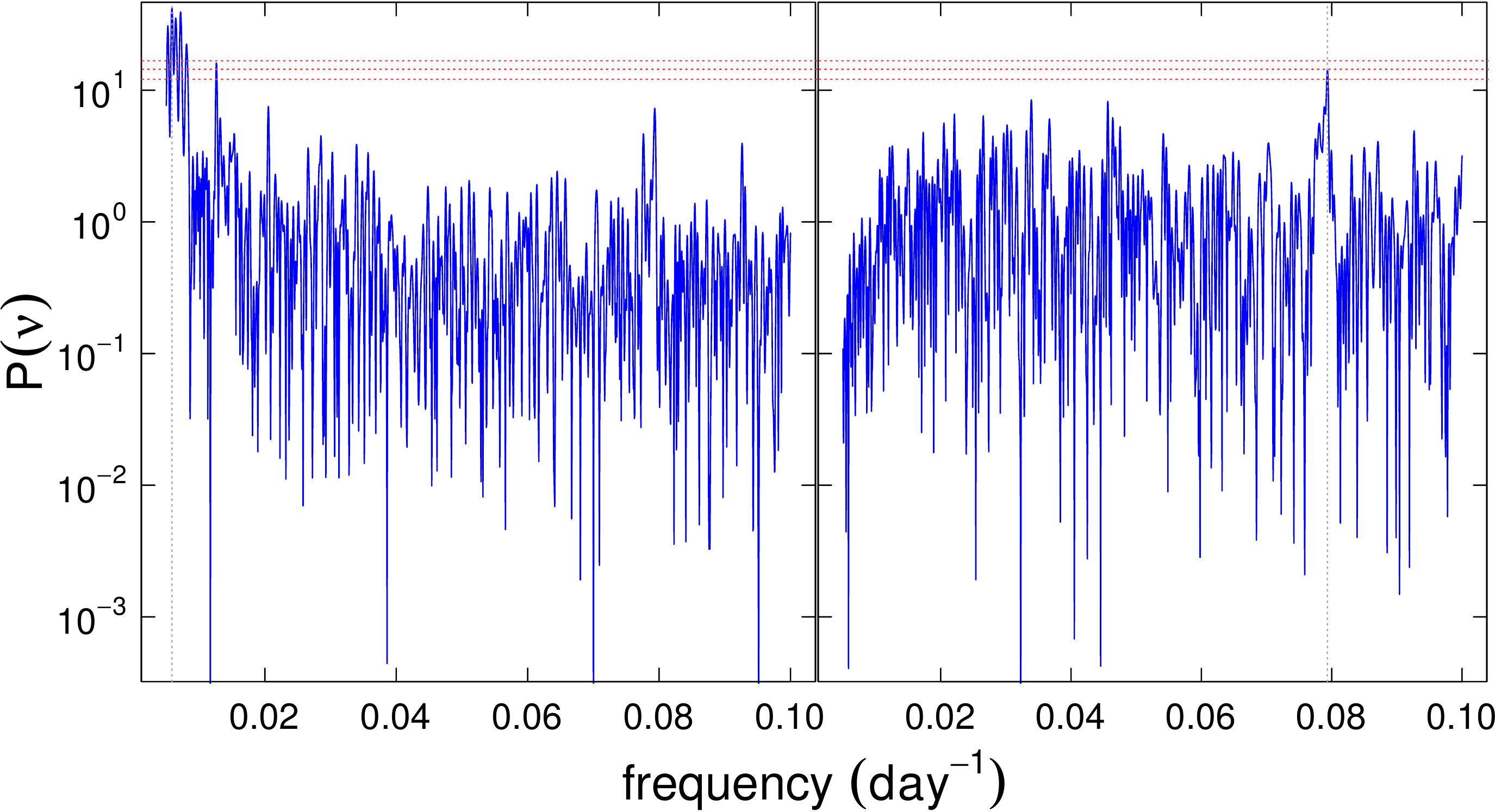}
\caption{
         Lomb--Scargle periodograms of our {\it Swift} data prior to (left panel)
         and after (right panel) removal of low-frequency signatures. Horizontal
         lines in both panels show the false-alarm levels of 1\%, 0.1\% and
         0.01\% (respectively from bottom to top). A prominent feature is seen at
         $\sim$\,0.079{\ts}days$^{-1}$ ($\sim${\ts}$12.61${\ts}days).	
         \label{fig04}
        }
\end{figure}

We performed a further analysis in the {\it Swift} database,
since its 0.5 days$^{-1}$ Nyquist frequency allows us to explore
periodic signatures as short as 2 days in period. The first step was to remove the
low-frequency modulations and trends from the long-term light curve. This procedure
(also known as {\it detrending}) is based in the LOcally
Weighted Scatterplot Smoothing (LOWESS) polynomial interpolation
\citep{lowess} and was applied with a span (i.e. smoothing parameter)
of 0.05. The result is presented in Figure{\ts}{\ref{fig04}}, which
shows the Lomb--Scargle periodograms prior to and after the detrend. A clear 
signal around 0.079{\ts}days$^{-1}$
($\sim${\ts}$12.61${\ts}days) emerged after the removal of the low-frequency modulations.

\begin{figure}[ht!]
\figurenum{5}
\epsscale{1.13}
\hspace{-0.4cm}
\plotone{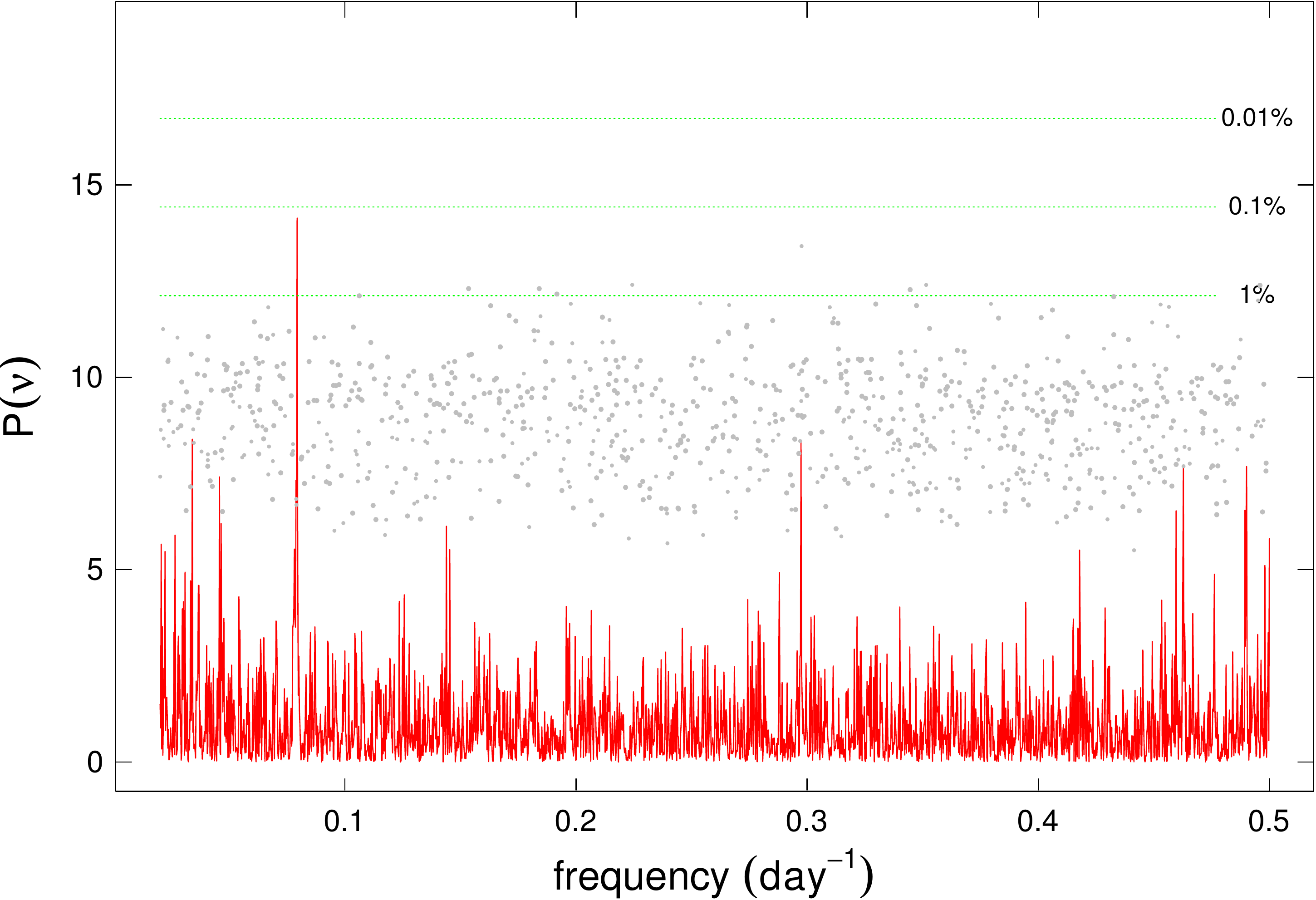}
\caption{
         The Lomb--Scargle periodogram of the {\it Swift} database (in red) together
         with peak values for the data generated by scrambling (in gray: see the text for details). We also show false-alarm levels. The periodic
         signature, interpreted as the orbital period of {\E}, is clearly seen at $\sim$\,0.079{\ts}days$^{-1}$. 
         \label{fig05}
        }
\end{figure}

In order to better constrain the significance of this apparent
periodicity in the {\it Swift} data, we applied a \textit{scrambling}
procedure, which is done by preserving the original time tags of the
database but interchanging pairs of fluxes. A new {\it scrambled}
light curve is produced when 95{\ts}{\%} of the flux values is
switched. We generated 1000 new datasets with this procedure and then
calculated the Lomb--Scargle periodograms for
each. Figure{\ts}{\ref{fig05}} shows the {\it maximum} values of the
periodogram for the 1000 scrambled data (gray dots), as well as the
periodogram for our original {\it Swift} database (red lines). It can
be seen that neither the false-alarm levels were surpassed nor a
concentration of scrambled data maxima occurred nearby the 12.61 day
period, thus considerably reducing the chance of the latter existing
due to noise or deficient sampling.

\begin{figure}[ht!]
\figurenum{6}
\epsscale{1.13}
\hspace{-0.4cm}
\plotone{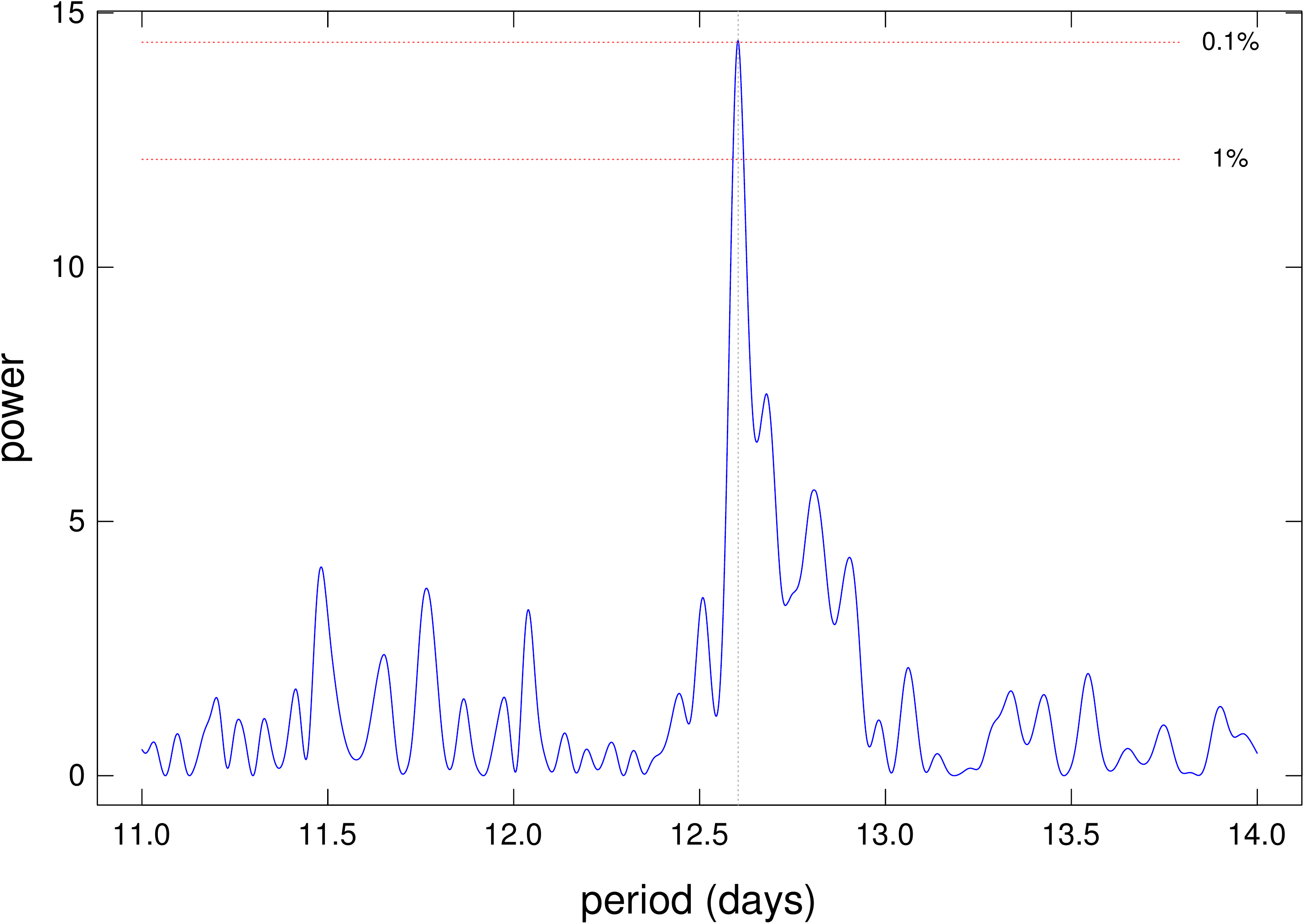}
\caption{
         The Lomb--Scargle periodogram for the {\it Swift} database around the 12.61 day modulation. 
         \label{fig06}
        }
\end{figure}

For a better visualization,
Figure{\ts}{\ref{fig06}} presents the power spectrum
with a magnification around the peak's region and with the horizontal axis
expressed as period. The peak value corresponds to $12.61${\ts}$\pm${\ts}$0.06${\ts}days. The uncertainty in the peak 
location was calculated using an expression very similar to that presented 
by \cite{1986ApJ...302..757H},

\begin{equation}
 \sigma_P = \frac{1}{2\pi}\sqrt{\frac{24}{N}}\frac{P^2}{T}\frac{\sigma}{R},
\end{equation}
where $N$ is the number of data points, $T$ is the span of the light
curve, $\sigma^2$ is the variance of the noise around the signal, and
$R$ is the semi-amplitude of the signal. However, this tends to
produce an optimistic estimate of $\sigma_P$ because it does not
take into account the specific timing characteristics (e.g., gaps
between measurements) of the light curve. A more realistic
value is obtained using the number of bins in a phase-folded diagram
for $N$, in which case the ratio $P^2/T$ becomes $\sim$\,1. By
calculating this way, $\sigma_P$ corresponds to
$\sim$\,0.06\,days -- value we adopted for the uncertainty in the
12.61-day signal.

\begin{figure}[ht!]
\figurenum{7}
\epsscale{1.13}
\hspace{-0.4cm}
\plotone{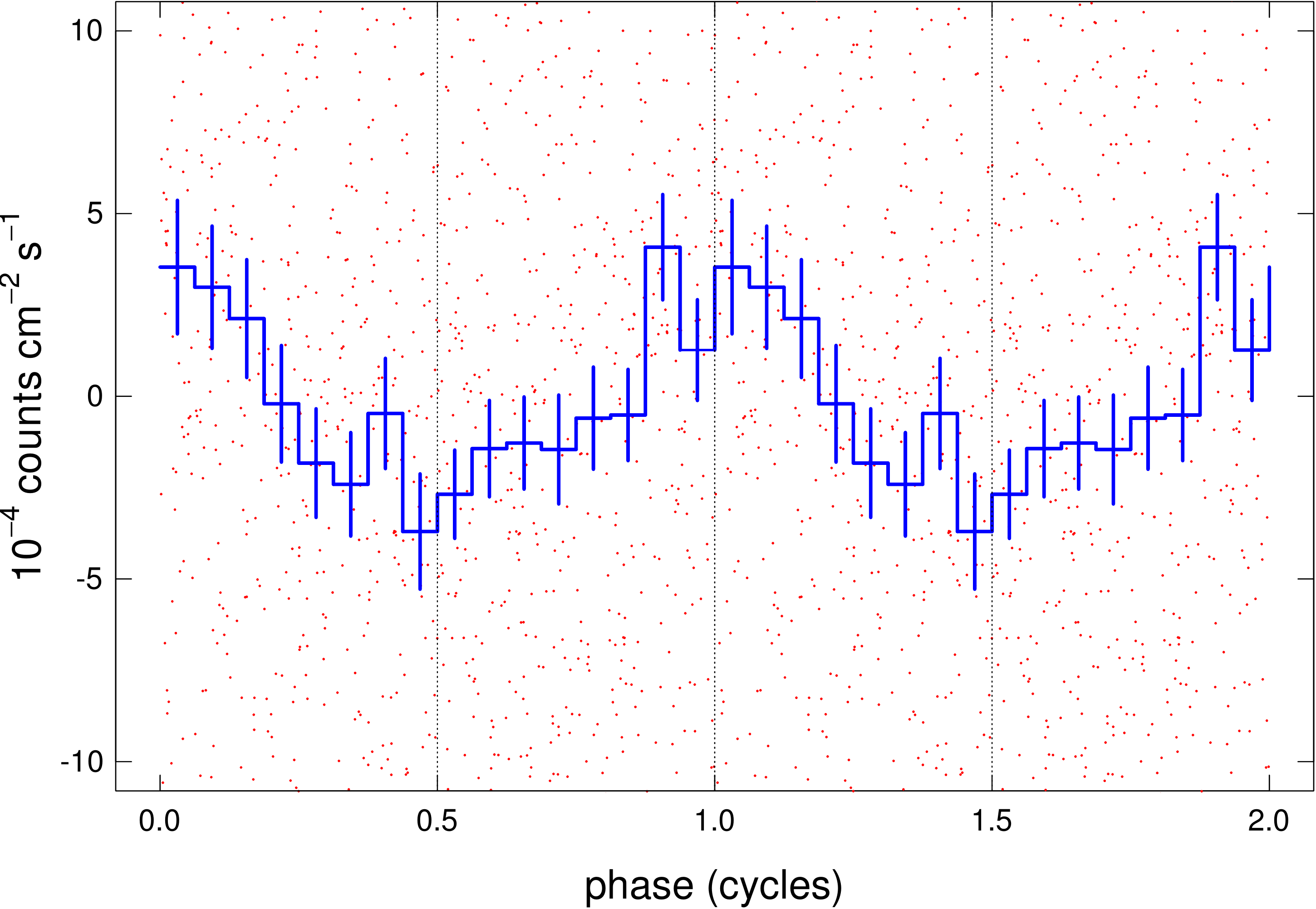}
\caption{Folded-phase light curve of {\E} for the orbital period
         ($12.61${\ts}days). Here, the scatter of the individual points is larger compared to Figure\,{\ref{fig03}} due to the smaller amplitude of this signal.         
         \label{fig07}
        }
\end{figure}

An ephemeris for the times of maximum of the 12.61 day signal can
be expressed as
MJD\,53424.85{\ts}$\pm${\ts}$E$\,$\times${\ts}$12.61${\ts}days. The folded-phase light curve for this period is shown in Figure{\ts}{\ref{fig07}}. Our interpretation is that this modulation is related to the orbital period of the system (see the discussion
below). Both the period and the semi-amplitude normalized to the average count rate (3.5\%) are consistent with the values previously measured in the \textit{RXTE} mission \citep{2002ApJ...578L.129S} -- $12.73${\ts}$\pm${\ts}$0.05${\ts}days and $\sim$\,3.4\%, respectively.

\section{Discussion}

The modulation at $171.1${\ts}$\pm${\ts}$3.0${\ts}days (shown in
Figure{\ts}{\ref{fig02}}) was never reported before for {\E}. In X-ray
binaries, long-term modulations, i.e., superorbital periods,
are interpreted as a precession of the accretion disk (see, e.g.,
\citealp{1974ApJ...187..575R}, and \citealp{1980ApJ...236..255M}).  If
this proves to be the case, it is expected that variations in this
superorbital period should occur on timescales of several years
\citep{1999A&A...343..861B}. This has been seen, for example, in
Cyg{\ts}X-1, where data available since the {\it Vela} and {\it Ariel}
missions up to the {\it RXTE} and {\it Swift} monitoring point to an
{\it increase} from $\sim${\ts}290 to $\sim${\ts}326{\ts}days of the
superorbital period \citep{2008ApJ...683L..55R}.  A continued
monitoring of {\E} would then be interesting to search for similar
variations. A previous report of a tentative superorbital period in
{\E} suggested a value around 600{\ts}days \citep{2002ApJ...578L.129S},
significantly different from the value reported here. However, our
procedure of examining simultaneously two independent datasets greatly
improves the reliability of detecting possible periodic signals. We thus
propose, by similarity to other systems, that the $171.1$\,day
feature is the superorbital period associated with the precession of the
accretion disk in the system.

Because {\E} is very likely a binary system, the periodic modulation at
$12.61${\ts}$\pm${\ts}$0.06${\ts}days could be due to the companion
star causing a partial eclipse of the region where the hard X-rays are
produced, which would occur at every orbit. Assuming that the hard
X-ray flux in {\E} is due to Comptonization of seed (disk) soft X-ray
photons by a corona of relativistic electrons, as proposed by some
authors (e.g., \citealp{cas14}), then one natural explanation of the
modulation is the occultation of such a corona by the companion.

Within 2{\ts}$\sigma$, this orbital period agrees with the value of
$12.73${\ts}$\pm${\ts}$0.05${\ts}days reported by
\cite{2002ApJ...578L.129S}, which made use of PCA/\textit{RXTE} data.
Furthermore, we note that the orbital periods for 4 HMXB with black
holes, Cyg{\ts}X-1, LMC{\ts}X-1, LMC{\ts}X-3, and LS{\ts}I{\ts}+61, are
already established to be between $\sim${\ts}2 and $\sim${\ts}26{\ts}
days, so our value is in the right order of magnitude for this kind of
system.

Data from the BlackCat Catalog
\citep{2016A&A...587A..61C} show that for 21 LMXB BH/BHCs the orbital periods are
firmly measured. A clear outlier in such a database is the source
GRS{\ts}1915$+$105 with an orbital period of $\sim$\,33.8{\ts}days. The values of mean, median, and standard deviation for the
remaining 20 sources are $1.1$, $0.3$, and
$1.6${\ts}days, respectively. If from this subset we further withdraw
V{\ts}404{\ts}Cyg, with an orbital period of 6.5{\ts}days, the numbers become (mean, median, and standard deviation, respectively): $0.8$,
$0.3$, and $0.9${\ts}days.  Such results, of course, point to a highly
skewed distribution for the measured orbital periods of LMXB BH/BHCs in
the direction of shorter (typically less than a day) values. It is thus tempting 
to assume that the $12.61$\,day
orbital period of {\E} points to HMXB nature for the binary system.
It is also worthwhile recalling that all reported orbital
periods for LMXB BH/BHCs were measured for {\it transient} sources,
whereas the known HMXB BH/BHCs for which orbital periods were measured are
{\it persistent} sources -- which is the case of {\E}. Therefore, it is our interpretation that 
{\E} is HMXB BHC, as recently suggested by studies in longer wavelengths
(e.g., \citealp{2015A&A...584A.122L}).

\acknowledgments

\textit{Acknowledgments.}  P.E.S. acknowledges CAPES/Brazil for
support. M.C. acknowledges financial support under grant
\#2015/25972-0, S\~ao Paulo Research Foundation (FAPESP) as part of
Thematic Project \#2013/26258-4 from the same foundation. J.B. also
acknowledges FAPESP under Thematic Project \#2013/26258-4.
We thank an anonymous referee for comments that helped us to improve
the quality of this study.



\end{document}